# Large Language Models in Student Assessment: Comparing ChatGPT and Human Graders

Magnus Lundgren*

Department of Political Science, University of Gothenburg

**Abstract**

*This study investigates the efficacy of large language models (LLMs) as tools for grading master-level student essays. Utilizing a sample of 60 essays in political science, the study compares the accuracy of grades suggested by the GPT-4 model with those awarded by university teachers. Results indicate that while GPT-4 aligns with human grading standards on mean scores, it exhibits a risk-averse grading pattern and its interrater reliability with human raters is low. Furthermore, modifications in the grading instructions (prompt engineering) do not significantly alter AI performance, suggesting that GPT-4 primarily assesses generic essay characteristics such as language quality rather than adapting to nuanced grading criteria. These findings contribute to the understanding of AI's potential and limitations in higher education, highlighting the need for further development to enhance its adaptability and sensitivity to specific educational assessment requirements.*

## Introduction

In the realm of higher education, grading remains a pivotal yet challenging task, directly impacting the quality of education and the workload of educators [1, 2]. As student populations grow and the range of assessments diversify [3], the traditional methods of grading, often labor-intensive and subject to human biases, call for innovative solutions. This study investigates the performance of Generative Pretrained Transformers (GPTs), a type of AI language model popularized by OpenAI, in assessing and grading student essays.

The central inquiry focuses on the efficacy and reliability of GPT-4 as a grading tool. Specifically, the study aims to answer the question of whether GPT-4 can provide accurate numerical grades for written essays in higher social science education. The study utilizes a sample of 60 anonymized master-level essays previously graded by teachers as a benchmark to assess the performance of GPT-4, using a variety of instructions ("prompts") to explore variation in quantitative measures of predictive performance and interrater reliability.

The study makes three critical findings regarding the use of GPTs for grading essays in higher social science education. First, GPT-4's grading closely aligns with human graders in terms of mean scores, though it exhibits a conservative grading pattern, primarily assigning grades within a narrower middle range. Second, GPT-4 demonstrates relatively low interrater reliability with human graders, as evidenced by a Cohen's kappa of 0.18 and a percent agreement of 35%, indicating significant room for improvement in AI grading alignment with human judgment. Third, the investigation reveals that adjustments to the grading instructions via prompt engineering do not significantly influence GPT-4's performance. This suggests that the AI predominantly evaluates essays based on generic characteristics such as language quality and structural coherence, rather than adapting to the detailed and nuanced assessment criteria embedded within different prompts.

The absence of a human-to-human comparison for the same set of essays limits our understanding of how GPT-4's interrater reliability stacks up against typical human variance in grading. However, this limitation notwithstanding, the study's empirical findings contribute to a growing literature on using AI for grading and evaluation in higher education [4, 5, 6, 7], suggesting three principal implications. First, although AI has the potential to develop into a resource-efficient assessment tool, reducing the grading workload for university teachers, further technological development is required to improve alignment with human raters. Second, the research underscores the challenge AI presently faces in grading complex, lengthy essay materials compared to simpler, more deterministic tasks like exam questions. Third, the consistent performance of GPT-4 across different prompts reveals a limitation in its ability to differentiate based on nuanced grading instructions, suggesting a risk of misclassification in cases where linguistic flair exceeds analytical quality.

---

*magnus.lundgren@gu.se

# AI and Language Models in Student Evaluation

The development of automated essay scoring (AES) systems has significantly evolved over the past decades. Initially conceptualized in the late 1960s with Project Essay Grader (PEG), AES technologies have grown in sophistication, incorporating advanced machine learning techniques to evaluate the quality of written texts based on features like essay length, word length, and syntactic variety (see [8] for a review). Following the arrival of LLMs, and specifically GPTs, an emerging body of literature has started to probe their performance and other characteristics in classifying and evaluating written text in the context of higher education, thereby moving beyond more conventional machine learning approaches (e.g., [9]).

One line of research has focused on the potential to employ GPTs in the service of research assistance tasks. A study by [4] showed that GPT-3 could replicate human-sourced survey responses while [5] found that LLMs could be used as surrogates for human subjects in certain types of social science research. Similarly, exploring whether this type of AI models can be utilized for research assistance tasks, such as automated content classification in social science research, Lupo et al. (2023: 1) found that "an LLM can be as good as or even better than a human annotator while being much faster".

Another literature focuses on the usage of AI in student assessment. [10] review a variety of applications of AI in higher education student assessment. They highlight a growing trend in the use of AI for formative assessment, providing immediate feedback and grading, and point to how AI has been used to assist teachers with large groups of students. At the same time, [10] observe a lack of widespread use of AI in education due to limited knowledge among users and the need for specific training for effective implementation.

Focusing on how AI tools can assist in feedback, [6] examine the efficacy and student preferences concerning AI-generated feedback in writing for English as a New Language (ENL) learners. They report two longitudinal studies: The first assesses if AI-generated feedback (using GPT-4) impacts learning outcomes compared to human tutor feedback, finding no significant difference between the two. The second study explores ENL students' preferences for AI-generated versus human tutor feedback, revealing no significant differences in preferences, while suggesting that both types of feedback may have distinct advantages. The authors conclude that a blended approach, integrating both AI and human feedback, could be most effective.

On the topic of using AI in grading, [11] provides a conceptual discussion, reflecting on potential benefits of AI in grading (such as efficiency and consistency) and drawbacks (such as privacy concerns and the quality of feedback). Anoter study [7] used ChatGPT (GPT-3) to grade open-ended questions answered by 42 industry professionals in technical training. The effectiveness of ChatGPT was evaluated by comparing its corrections and feedback against subject matter experts. The results suggested that the AI had some capability to identify nuanced semantic details and that subject matter experts "tended to agree with the corrections and feedback given by ChatGPT".

These studies represent significant advancements in our understanding of AI tools in higher education assessment but also exhibit limitations. Most importantly, they have remained focused on using AI in the grading of shorter exam questions. The challenges in grading exam questions are likely different from those that emerge from grading student essays, which are less focused on providing a factual answer and typically require—and assess—a greater range of analytical skills. Exploring whether and when LLMs have ability to grade essay assignments is therefore an important line of inquiry. Another limitation in existing research is their reliance on LLMs of inferior capability. Most published studies are carried out on early versions of GPTs, predominantly GPT-2 and -3, motivating further studies of more recent and capable versions of these models.

# Methodology

The study was performed using OpenAI's GPT-4 model, available via API[1] and the ChatGPT web interface.[2] At the time of the research, this model ranked as the most capable GPT publicly available.

The validation set consists of a sample of 60 student essays from a master-level course in political science offered at the University of Gothenburg in 2022 and 2023. The essay assignment requires students to write an individual research paper analyzing the impact of a specific policy, international agreement, or intervention on a particular aspect of an assigned country's social, political, institutional, or economic conditions, focusing on detailed empirical analysis and guided by academic literature. The essays were around 3,500 words, structured into 4-6 subsections, and typically included at least one or two figures or graphs.

Human raters graded the essays on a numerical scale, 1-7, taking into consideration clarity and struc-ture, integration of academic literature, empirical rigor, and quality of writing and presentation, with higher scores indicating clearer organization, deeper analysis, better integration of academic sources, more rigorous empirical work, and superior writing quality. All human grades were motivated with attendant written comments.[3]

---
[1] https://platform.openai.com/docs/api-reference
[2] https://chat.openai.com

This validation set is free from "data contamination" which would occur if the coded material was part of the LLM's training material [12]. The temporal cutoff of the used GPT-4 model preceded the grading and the essays exist in the public domain.

Ethical considerations were paramount in this study, with a strong focus on data privacy. Ensuring the confidentiality of student data was critical; hence, the study implemented rigorous anonymization protocols to prevent any possibility of identifying individuals from the essay data. These protocols were designed to comply with the General Data Protection Regulation (GDPR) and institutional guidelines. All papers were anonymized and any information that could tie the text to a person or institution was removed prior to handling by ChatGPT. Furthermore, prior to the execution of the study, all GPT settings were modified to minimize privacy concerns, including opting out from sharing conversations with OpenAI to train future GPT models.[4] Additionally, all tests were run in a zero-shot setup, meaning that no two grading processes influenced each other.

To assess the overall viability of the approach and to calibrate the research protocol, a pilot study was carried out (reported in Appendix A).

The main study was carried out according to the following protocol:

1. Input of individual essay in pdf format;
2. Input of grading instructions ("prompt");
3. Collection of output data (suggested grade).

Steps 1-3 were repeated for each of the 60 essays using the same prompt, yielding 60 observations (suggested grades) per prompt.

To probe GPT-4's sensitivity to different instructions and settings, the protocol was repeated with four different prompts (full information is available in Appendix B). The (1) basic prompt requests grading a master's level political science paper on a 1-7 scale based purely on the submitted paper. The (2) grading criteria prompt added a specific grading matrix for grading the paper, stipulating more specific criteria for a set of dimensions, each graded 1-7. The (3) reference prompt introduced two reference papers to calibrate the grading, providing examples of mid-range and high-grade standards. The (4) everything prompt combined elements from the previous ones—reference papers and a grading matrix—and additionally specifies an expected grade distribution, making it the most complex and structured grading instruction. The expectation was that, if GPT-4 can adapt to more detailed and granular instructions, these modified prompts would improve its ability to predict grades set by human raters.

All prompts were submitted in English, corresponding to both the course language and the language of the student assignments. Recent research suggests that GPT-4 attains a higher performance in coding material in the English language compared with other languages [13].

The collected data were evaluated for interrater reliability and overall predictive accuracy. Following [14], interrater reliability was assessed based on both consensus and consistency estimates. Consensus is assessed based on percent agreement, calculated by dividing the number of GPT-human agreement by the total number of observations, and Cohen's kappa, which adjusts percent agreement for the agreement that could happen by chance. Consistency is evaluated based on the Pearson correlation coefficient, which is the most conventional correlation measure, and Spearman's rank correlation coefficient, which measures how well two sets of ranks correlate. Predictive accuracy was also evaluated based on ordinary least squares (OLS) regression.

## Results

Descriptive and distributional statistics (Table 1, Figure 1) suggest that the mean GPT-4 grades align with that of human raters. The GPT-4 rater with the basic prompt grades somewhat more leniently, but the other GPT-4 raters arrive at grade distributions with means that are statistically indistinguishable from that of human raters. Inspection of the distributions (Figure 1) suggests that the GPT-4 raters produce grade distributions with a narrower range, which is also confirmed by the observed standard deviations, which are considerably smaller for GPT-4 raters than human raters. Overall, these measures indicate that GPT-4 raters exhibit a "bias towards the middle," awarding fewer essay grades at the lower and higher ends of the spectrum compared with human raters. Taken together, this suggests that GPT-4 is risk averse, seemingly avoiding assigning low or high grades. Potentially, this is suggestive of the "polite" or "optimistic" traits that researchers have attributed to GPTs in other studies [15].

In terms of interrater reliability (Table 2), GPT-4 raters exhibit relatively low levels of agreement

---

[3] This study focuses solely on the numerical grades. Further research should investigate whether LLMs can generate written comments that align with those provided by human raters.
[4] https://privacy.openai.com/policies

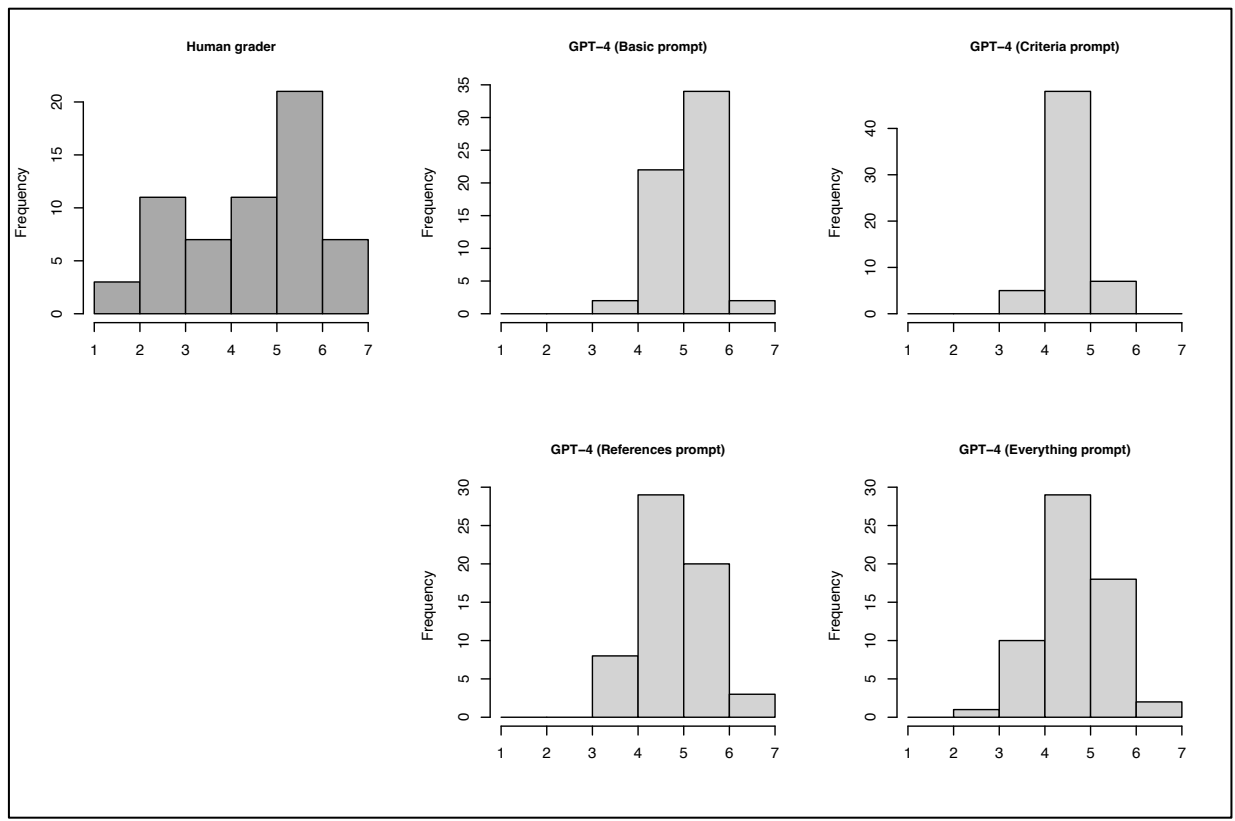

**Figure 1:** *Distribution of essay grades: human raters and GPT-4*

**Table 1:** *Essay grades descriptive statistics*

|                    | Mean   | S.D. |
|--------------------|--------|------|
| Human raters       | 4.95   | 1.47 |
| GPT-4: Basic       | 5.60*  | 0.62 |
| GPT-4: Criteria    | 5.03   | 0.45 |
| GPT-4: References  | 5.30   | 0.77 |
| GPT-4: Everything  | 5.17   | 0.81 |

* t-test indicates difference in means at p<0.01 compared with human raters.

with human raters. The best-performing GPT-4 rater overall is the basic prompt, which attains a 35 percent agreement and a Cohen's kappa of 0.18, whereas other prompts tend to have lower scores. Since we lack a human-to-human comparison for this essay set, we cannot compare it to human-level interrater reliability for this exact material. However, these are interrater reliability levels that would be interpreted as low under conventional circumstances.

Given the nature of the essay assignment and the grading system, it is motivated to place more emphasis on the two consistency measures, which are less sensitive to exact matches. These measures suggest that there is an overall correlation between human and GPT-4 grades, suggesting that human and GPT raters attribute similar importance to the core elements of the essay content. This also implies that while GPT-4 and human raters may prioritize similar criteria when assessing essays, the interpretation of these criteria into numerical grades exhibits variability. These results are largely replicated by OLS regressions, which also provide additional information on the uncertainty of estimates (see also Figure 2).

While we may want to place more emphasis on the two consistency measures, the results nevertheless indicate that prompt engineering does not significantly alter the performance beyond the basic prompt. This suggests that GPT-4 predominantly assesses generic features of essay content, such as the quality of the language or style of writing, rather than adapting to the nuanced requirements embedded within various prompt configurations.

## Conclusion

This study has evaluated the ability of using LLMs, specifically OpenAI's GPT-4 model, to grade master-level student essays in social science. A sample of 60 essays was processed through four different grading prompts to assess how well the AI's suggested grades aligned with those given by human graders. Empiri-cal findings revealed that GPT-4 closely matches human grading standards in terms of mean scores but exhibits a bias towards middle-range grades, resulting in a risk-averse assessment profile. Importantly, the AI demonstrated relatively low interrater reliability with human raters and adjustments in grading instructions did not significantly influence its performance, suggesting that GPT-4 evaluates based on generic features of the essays rather than specific grading criteria.

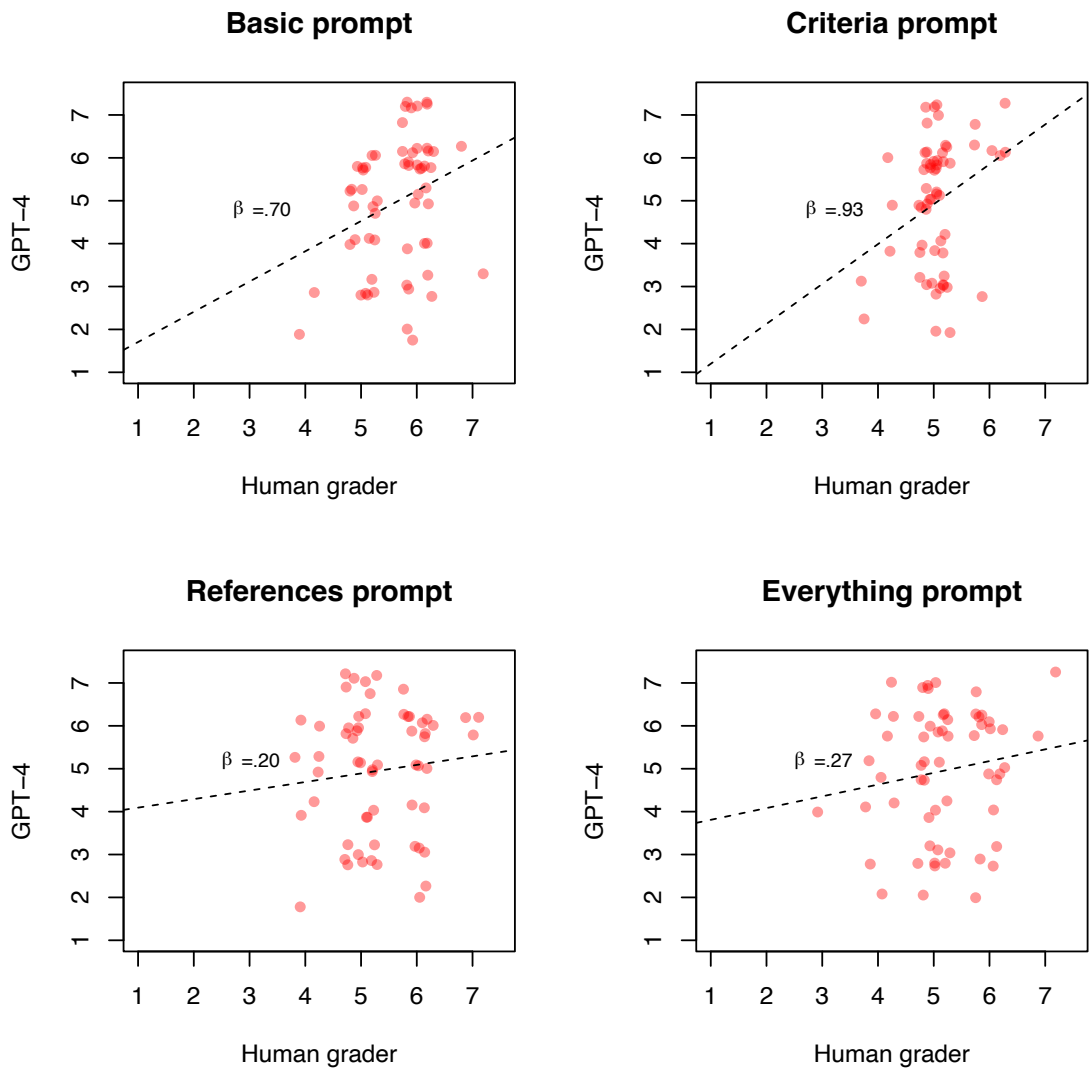

**Figure 2:** *Scatter plots of essay grades: human raters vs. GPT-4*

|                        | Percent agree | Cohen's kappa | Pearson's correlation | Spearman's rho |
|------------------------|---------------|---------------|-----------------------|----------------|
| GPT-4: Basic           | 35            | 0.18          | 0.30                  | 0.32           |
| GPT-4: Criteria        | 25            | 0.16          | 0.31                  | 0.30           |
| GPT-4: Reference papers| 27            | 0.08          | 0.43                  | 0.10           |
| GPT-4: Comprehensive   | 27            | 0.13          | 0.26                  | 0.13           |

**Table 2:** *Interrater reliability scores by prompt. Note: Comparison is with human raters.*

In the absence of comparable measures of human interrater ability for the same material, it is difficult to assess the full meaning of these results. It is possible that reliability is comparable to what we would observe in a human-to-human comparison. However, the results appear to suggest three principal implications.

First, the interrater reliability scores, as indicated by Cohen's kappa and percent agreement scores, raise concerns about the consistency between human and AI grading of this type of essay material. Moreover, the narrower grade distributions and the overall lower interrater reliability with human raters suggest areas where further training of the AI models might be necessary to better mimic human grading behaviors and improve their utility in educational settings (and compliance with relevant regulations). Overall, the discrepancy suggests a need for better alignment of LLM grading mechanisms with human judgment patterns.

Second, the performance observed here is worse than LLM performance on related tasks such as grading of exam questions [7] and coding of political text [4]. This deviation is likely explained by differences in the coded material. For instance, whereas [7] concerned the grading of shorter and more deterministic exam questions and [4] asked LLMs to code shorter texts into topical categories, this study concerned the more difficult task of categorizing longer texts into numerically ordered categories. If so, it suggests that the complexity of the source material is an important source of variation in the expected performance of LLMs in grading in higher education.

Third, the stability across different prompt conditions highlights a limitation in LLM capacity to discriminate between the differences dictated by grading instructions. This is perhaps the most troublesome finding, as it suggests possible limitations to the calibration of LLMs to rate material based on grading matrices and similar instructions. If LLMs mainly pick up general features of the written text, such as language usage and style, it increases the risk of misclassification. Even if such general language features correlate with essay quality—as is likely often the case—this risk is non-negligible. For example, if a given essay is exceptionally creative with regard to its handling of empirical material but below average in terms of writing quality and style, a human rater may presently be more likely to account for the creative qualities, whereas LLMs appear more likely to overlook them.

In light of these findings—and the limitations they suggest—the overall conclusion is that LLMs are not yet a mature tool for assessment of longer essays in higher education. At the same time, we must assume that the AI technology will continue to evolve, providing teachers with gradually more sophisticated and tailored software solutions which may address several or all of the observed limitations. This would involve, e.g., sensitivity to specialized grading rubrics and prompt-specific criteria, greater ability to handle varied text and open-ended assignments, and ability to analyze the quality of arguments. As AI attains such capabilities, it likely to become a tool that teachers employ for a variety of tasks, including grading.

The introduction of AI in grading in higher education would raise several broader questions. One question pertains to the role of assessment plays for teachers in evaluating their students' learning. If grading is outsourced to an automatized agent, will teachers be as able to observe whether their students "are getting it"—and to adjust their course content accordingly—as they are when they do the grading themselves? Likely, the AI tools of tomorrow can be adjusted to provide the teacher with information about where a particular group of students fell short and where they did well, but it is not certain that such information can replace the experience of having "close contact" with the student essays themselves.

Likewise, if teachers are seen as free-riding on AIs in grading or unable to explain AI-determined grades with sufficient clarity, it may undermine teachers' legitimacy and students' motivations to learn [16].

Another question pertains to student assessment as an exercise of public authority (in many national systems). When AI systems are employed for grading in universities, defining legal accountability and ensuring transparency becomes crucial. These systems blur traditional lines of responsibility, raising concerns about who is accountable for AI decisions—whether it be the developers, the institution, or the individual educators using the technology. Moreover, the opaque nature of AI algorithms can conflict with the required transparency of public decisions, making it difficult for students to understand and challenge their grades. Practical integration of AI in student assessment would therefore depend on establishing regulatory frameworks that define accountability for AI-generated outcomes and creating mechanisms to ensure that AI decision-making processes in educational settings are as transparent and explainable as possible.

A final, more philosophical question arises when considering the role of AI in grading: Is anything significant lost when the assessment interaction between teacher and student is mediated by AI? In the pedagogy of higher education, grading is not merely a measure of student performance but also a critical feedback mechanism that influences learning strategies and outcomes. Some may argue that the personal insights and nuanced understanding that educators bring to the grading process could be diminished when replaced by AI. However, arguably, the fundamental aspects of grading—evaluating knowledge, comprehension, and critical thinking—can be efficiently mimicked by sophisticated algorithms. If AI continues to evolve to address its current limitations and social and regulatory norms evolve to permit its application in higher education, the technology may be able to enhance grading by providing more consistent and unbiased assessments. By reliably handling routine grading tasks, AI could then allow educators to dedicate more time to personalized teaching and mentoring [6, 11]. Thus, rather than detracting from human interaction, a more advanced AI could support and enhance educational engagement by freeing up human capacities for the deeper, more meaningful interactions between teacher and student than what is provided by conventional grading.

# References

[1] D. Allen, ed. *Assessing student learning: From grading to understanding*. Teachers College Press, 1998.

[2] C. Golding and L. Adam. "Evaluate to improve: useful approaches to student evaluation". In: *Assessment & Evaluation in Higher Education* 41.1 (2016), pp. 1–14.

[3] A. Bryntesson and M. Börjesson. *Breddad rekrytering till högre utbildning: En beskrivning och jämförelse av policy och utfall i de skandinaviska länderna*. Uppsala universitet, Humanistisk-samhällsvetenskapliga vetenskapsområdet, Fakulteten för utbildningsvetenskaper, Institutionen för pedagogik, didaktik och utbildningsstudier, 2021.

[4] L. P. Argyle et al. "Out of one, many: Using language models to simulate human samples". In: *Political Analysis* 31.3 (2023), pp. 337–351.

[5] G. V. Aher, R. I. Arriaga, and A. T. Kalai. "Using large language models to simulate multiple humans and replicate human subject studies". In: *International Conference on Machine Learning*. PMLR. 2023, pp. 337–371.

[6] J. Escalante, A. Pack, and A. Barrett. "AI-generated feedback on writing: insights into efficacy and ENL student preference". In: *International Journal of Educational Technology in Higher Education* 20.1 (2023), p. 57.

[7] G. Pinto et al. "Large language models for education: Grading open-ended questions using ChatGPT". In: *Proceedings of the XXXVII Brazilian Symposium on Software Engineering*. 2023, pp. 293–302.

[8] K. Yang et al. "Unveiling the Tapestry of Automated Essay Scoring: A Comprehensive Investigation of Accuracy, Fairness, and Generalizability". In: *Proceedings of the AAAI Conference on Artificial Intelligence*. Vol. 38. 20. 2024, pp. 22466–22474.

[9] Ken Benoit et al. "Text as data: an overview". In: *The SAGE handbook of research methods in political science and international relations*. 2020, pp. 461–497.

[10] V. González-Calatayud, P. Prendes-Espinosa, and R. Roig-Vila. "Artificial intelligence for student assessment: A systematic review". In: *Applied Sciences* 11.12 (2021), p. 5467.

[11] R. Kumar. "Faculty members' use of artificial intelligence to grade student papers: a case of implications". In: *International Journal for Educational Integrity* 19.1 (2023), pp. 1–10.

[12] Shahriar Golchin and Mihai Surdeanu. "Time travel in llms: tracing data contamination in large language models". In: *arXiv preprint arXiv:2308.08493* (2023).

[13] Kabir Ahuja et al. "Mega: multilingual evaluation of generative ai". In: *arXiv preprint arXiv:2303.12528* (2023).

[14] S. E. Stemler. "A Comparison of Consensus, Consistency, and Measurement Approaches to Estimating Interrater Reliability". In: *Practical Assessment, Research, and Evaluation* 9.1 (2004), p. 4.

[15] X. Wang et al. "Does role-playing chatbots capture the character personalities? assessing personality traits for role-playing chatbots". In: *arXiv preprint arXiv:2310.17976* (2023).

[16] K. Struyven, F. Dochy, and S. Janssens. "Students' perceptions about evaluation and assessment in higher education: A review". In: *Assessment & evaluation in higher education* 30.4 (2005), pp. 325–341.

# Appendix A: Pilot Study

Prior to the full study, a pilot study was carried out. This involved the grading of a set of three papers (called papers A, B, and C below). The purpose of the pilot study was to assess the general viability of the approach and how variation in prompting affected results. No statistical analysis was carried out.

The papers were written as the final research assignment within the course "International Administration and Policy" and examine various policy interventions by international institutions in specific developing countries. Each paper was around 3500 words, excluding references.

Procedure: Each paper was submitted to ChatGPT (GPT-4) together with prompts of gradually increasing specificity of instruction (Table 3). Each iteration involved uploading the paper, possible reference material, and inputting the prompt. The suggested numerical grade was collected from GPT-4 output (Table 3). Comments volunteered by GPT-4 were not collected.

At that stage, three preliminary impressions were noted:

First, while N is very low and no statistical analysis has been performed, the results suggest the possibility of alignment between GPT-4 and human graders. The average GPT grades suggest an overall ranking of papers A, B, and C equivalent to that implied by the human grader.

Table 3: *Summary of results for papers A-C*

| Grader | A | B | C |
|---|---|---|---|
| GPT-4: Basic | 6 | 6 | 6 |
| GPT-4: Basic + grading criteria | 5.5 | 6 | 6 |
| GPT-4: Basic + reference paper | 5.5 | 4 | 6 |
| GPT-4: Basic + grading criteria + reference paper | 6 | 5 | 6 |
| Human grader | 5 | 4 | 7 |

Grades for papers A, B, and C.

Second, GPT-4 is biased upwards for papers A and B and downwards for paper C. Taken together, this suggests that GPT-4 is risk averse, seemingly avoiding assigning low or high grades. Potentially, this is suggestive of the "polite" or "optimistic" traits that researchers have attributed to GPT-4.

Third, increasing the granularity and specificity of prompts appears to have the potential to increase alignment between GPT-4 and human graders. In this small sample, the provision of a reference paper with a pre-assigned grade generated greater alignment.

# Appendix B: Prompts

**Basic:** I want you to act as a teaching assistant grading a student paper from a master's level course in political science. I will input a paper and I want you to grade the paper on a scale from 1 to 7 where 1 is a failing grade and 7 the highest possible grade. Grade the attached paper and provide the numerical grade (nothing else).

**Grading criteria:** I want you to act as a teaching assistant grading a student paper from a master's level course in political science. I will input a paper and I want you to grade the paper on a scale from 1 to 7 where 1 is a failing grade and 7 the highest possible grade. Grade the attached paper using the also attached grading matrix ("grading matrix"). Provide the numerical grade and nothing else.

**Reference paper:** I want you to act as a teaching assistant grading a student paper from a master's level course in political science. I will input a paper and I want you to grade the paper on a scale from 1 to 7 where 1 is a failing grade and 7 the highest possible grade. I will attach two reference papers, the first of which received a grade of 4 ("reference 4") and the second of which received a grade of 7 ("reference 7"). Grade the attached paper ("paper") calibrating against the quality of the reference papers. Provide the numerical grade and nothing else.

**Everything:** I want you to act as a teaching assistant grading a student paper from a master's level course in political science. I will input a paper and I want you to grade the paper on a scale from 1 to 7 where 1 is a failing grade and 7 the highest possible grade. I will attach two reference papers, the first of which received a grade of 4 ("reference 4") and the second of which received a grade of 7 ("reference 7"). I will also attach a grading matrix ("grading matrix") that show the requirements for each grade. Grades should average around 5 with a standard deviation of 1.5. Grade the attached paper ("paper") calibrating against the quality of the reference papers and taking the grading matrix and overall grade distribution into account. Don't be afraid to award low or high grades for papers that are clearly better or worse than the average. Provide the numerical grade and nothing else.

# Appendix C: Grading Matrix

**Clarity and structure**

1. Paper lacks logical flow; sections are disjointed and the introduction, body, and conclusion are unclear or missing.
2. Paper lacks logical flow; sections are disjointed and the introduction, body, and conclusion are unclear or missing.
3. Basic structure is present, but transitions between sections are weak; organization is somewhat confusing.
4. Adequate structure; clear sections, but some transitions might be abrupt.
5. Well-organized; good flow between sections and clear overall structure.
6. Very well-organized; sections are logically structured and transitions are smooth, enhancing readability.
7. Exceptionally clear and structured; the organization enhances the argument's effectiveness and engages the reader throughout.

**Analysis**

1. Superficial analysis with major misunderstandings of the topic; lacks critical engagement with data.
2. Superficial analysis with major misunderstandings of the topic; lacks critical engagement with data.
3. Some analysis present but lacks depth; minimal critical engagement with sources.
4. Satisfactory analysis with reasonable interpretation of data; demonstrates an understanding of the topic.
5. Good, detailed analysis; uses data effectively to support arguments.
6. Very strong analysis; demonstrates depth and insight in data interpretation and critical engagement.
7. Outstanding depth in analysis; insightful and demonstrates sophisticated understanding and application of data.

**Integration of literature**

1. Inadequate or incorrect use of literature; major citation issues.
2. Inadequate or incorrect use of literature; major citation issues.
3. Uses some relevant literature but often relies on non-academic sources or superficial references.
4. Good use of academic sources but integration into the argument could be improved.
5. Strong use and integration of academic literature to support the paper's arguments.
6. Very strong integration, using a diverse range of relevant academic resources effectively.
7. Exceptional integration of literature; uses academic sources creatively and effectively to advance a compelling argument.

**Empirical rigor**

1. Empirical content is largely inaccurate or irrelevant; sources are unreliable or misinterpreted.
2. Empirical content is largely inaccurate or irrelevant; sources are unreliable or misinterpreted.
3. Some relevant empirical content but with inaccuracies or generalizations.
4. Reasonable accuracy and relevance of empirical content with minor errors.
5. Accurate and relevant empirical analysis, well-supported by reliable sources.
6. Very accurate empirical work, detailed and well-supported by high-quality sources.
7. Exceptionally rigorous and accurate empirical analysis, setting a high standard for academic research.

**Quality of writing**

1. Numerous grammatical, stylistic, or formatting errors; does not follow academic writing standards.
2. Numerous grammatical, stylistic, or formatting errors; does not follow academic writing standards.
3. Some errors in grammar and style; inconsistent adherence to academic writing stan-

dards.
4. Generally follows academic writing standards with minor errors.
5. Good quality of writing; well-edited and formatted correctly according to academic standards.
6. Very high-quality writing; very few errors and excellent adherence to academic style.
7. Exceptional writing quality; flawless in terms of grammar, style, and adherence to the highest academic standards.